# Análise Semântica Automatizada com LLM e RAG para Bulas Farmacêuticas

**Daniel Meireles do Rego[1]**

[1]Instituto de Biociências, Letras e Ciências Exatas – Universidade Estadual Paulista (UNESP)
São José do Rio Preto – SP – Brasil

`daniel.meireles@unesp.br`

***Abstract.*** *The production of digital documents has been growing rapidly in academic, business, and health environments, presenting new challenges in the efficient extraction and analysis of unstructured information. This work investigates the use of RAG (Retrieval-Augmented Generation) architectures combined with Large-Scale Language Models (LLMs) to automate the analysis of documents in PDF format. The proposal integrates vector search techniques by embeddings, semantic data extraction and generation of contextualized natural language responses. To validate the approach, we conducted experiments with drug package inserts extracted from official public sources. The semantic queries applied were evaluated by metrics such as accuracy, completeness, response speed and consistency. The results indicate that the combination of RAG with LLMs offers significant gains in intelligent information retrieval and interpretation of unstructured technical texts.*

***Resumo.*** *A produção de documentos digitais vem crescendo de forma acelerada em ambientes acadêmicos, empresariais e de saúde, apresentando novos desafios na extração e análise eficiente de informações não estruturadas. Este trabalho investiga o uso de arquiteturas RAG (Retrieval-Augmented Generation) combinadas com Modelos de Linguagem de Grande Escala (LLMs) para automatizar a análise de documentos em formato PDF. A proposta integra técnicas de busca vetorial por embeddings, extração semântica de dados e geração de respostas em linguagem natural, contextualizadas. Para validar a abordagem, realizamos experimentos com bulas de medicamentos extraídas de fontes públicas oficiais. As consultas semânticas aplicadas foram avaliadas por métricas como precisão, completude, velocidade de resposta e consistência. Os resultados indicam que a combinação de RAG com LLMs oferece ganhos significativos na recuperação inteligente de informações e na interpretação de textos técnicos não estruturados.*

## 1. Introdução

A crescente digitalização de processos administrativos, acadêmicos, corporativos e de saúde tem gerado uma quantidade enorme de documentos eletrônicos, especialmente no formato PDF. Apesar de esses arquivos serem ricos em informações, suas estruturas textuais costumam variar bastante, com pouca padronização, o que torna a extração e análise eficiente de dados uma tarefa difícil para métodos tradicionais de busca por palavras-chave. Por isso, é fundamental buscar soluções que vão além de localizar palavras isoladas. É preciso interpretar conteúdos complexos de maneira contextualizada e semântica.

Nesse contexto, a Inteligência Artificial (IA) tem se consolidado como uma aliada importante para lidar com a complexidade dos dados atuais. Segundo Russell e Norvig (2021), a IA é uma área da ciência da computação dedicada ao desenvolvimento de sistemas capazes de realizar tarefas que, antes, dependiam exclusivamente da inteligência humana, como percepção, raciocínio, aprendizado e compreensão da linguagem.

Dentro da IA, o Processamento de Linguagem Natural (PLN) se destaca por possibilitar às máquinas entenderem, interpretarem e criarem textos em linguagem natural. Nos últimos anos, avanços em Modelos de Linguagem de Grande Escala (*LLMs — Large Language Models*), como o GPT-4 e o Gemini do Google, transformaram o PLN, ao permitirem interpretações e produções de textos com alta coerência semântica e forte contexto. Esses modelos, baseados em técnicas de aprendizado profundo, utilizam recursos como *embeddings* semânticos e mecanismos de atenção para captar detalhes sutis de significado em textos complexos.

Outro avanço importante vem das arquiteturas *Retrieval-Augmented Generation* (RAG), que combinam mecanismos de recuperação semântica de informações com geração de respostas em linguagem natural. Essa combinação de busca vetorial e geração assistida por contexto relevante traz mais precisão às respostas e ajuda a reduzir fenômenos como a "alucinação" de informações, comum em LLMs que funcionam sem suporte externo.

Essa integração entre recuperação e geração é especialmente útil na análise automatizada de grandes volumes de textos técnicos não estruturados, como documentos de alta complexidade.

Este trabalho investiga a aplicação de LLMs e arquiteturas RAG na análise automatizada de documentos digitais, usando bulas de medicamentos como estudo de caso. As bulas são documentos técnicos essenciais, contendo informações críticas sobre indicações, contraindicações, posologia e efeitos adversos. Por isso, representam um cenário ideal para validar estratégias de recuperação semântica e geração contextualizada.

A metodologia proposta combina técnicas de extração semântica, indexação vetorial com FAISS e geração de respostas fundamentadas por LLMs. Para testar a abordagem, realizamos experimentos com bulas de medicamentos reais, realizando consultas em linguagem natural e avaliando as respostas geradas com métricas de precisão, completude, tempo de resposta e consistência.

Este artigo busca não apenas apresentar a base teórica e a arquitetura da solução, mas também discutir os resultados obtidos, refletindo sobre as potencialidades e limitações do uso de LLMs combinados com arquiteturas RAG na automação da análise de documentos digitais em domínios técnicos específicos.

Além do contexto tecnológico e acadêmico, a proposta de aplicação também conta com respaldo legal. Segundo a Resolução RDC nº 47/2009 da Anvisa, as bulas de medicamentos devem ser formuladas de maneira clara, objetiva e acessível, facilitando a compreensão tanto por profissionais quanto por pacientes. A automação do acesso às informações, usando modelos de linguagem e consultas em linguagem natural, pode contribuir de forma efetiva para alcançar esse objetivo, ampliando o alcance e a compreensão do conteúdo técnico na área de saúde. Além disso, ao proporcionar uma alternativa de consulta baseada em documentos públicos e sem a necessidade de manipulação de dados pessoais, o sistema garante conformidade com os princípios da Lei Geral de Proteção de Dados (Lei nº 13.709/2018). A proposta também reforça o compromisso com a transparência e o direito à informação em saúde, conforme recomendações de diversas normativas do Sistema Único de Saúde.

## 2. Trabalhos Relacionados

A aplicação de inteligência artificial, especialmente os Modelos de Linguagem em Grande Escala (LLMs), vem se destacando na automação do entendimento e análise de documentos técnicos. Com o avanço de arquiteturas como GPT-4 (Brown et al., 2020) e Gemini (Google, 2023), tornou-se possível interpretar e gerar texto em linguagem natural com alto nível de coerência e contexto, impulsionando melhorias em tarefas como sumarização, respostas a perguntas específicas e extração de informações detalhadas.

Um grande desafio enfrentado pelos LLMs isoladamente é o fenômeno conhecido como alucinação, ou seja, a geração de informações plausíveis, mas incorretas ou infundadas. A arquitetura Retrieval-Augmented Generation (RAG), proposta por Lewis et al. (2020), surge como uma alternativa robusta ao permitir que os modelos consultem uma base de conhecimento externa por meio de buscas semânticas, antes de gerar uma resposta. Isso aumenta a confiabilidade, a precisão e a rastreabilidade das informações.

Diversos estudos têm validado essa abordagem. Pereira (2024) criou uma prova de conceito que integrou LLMs com RAG para analisar documentos jurídicos do Tribunal de Contas do Estado do Rio Grande do Sul. O sistema conseguiu automatizar a identificação de possíveis irregularidades em relatórios de auditoria, apresentando resultados promissores em termos de acurácia e agilidade. O autor também destacou o valor da arquitetura *ConversationalRetrievalChain* para criar uma experiência mais interativa para o usuário.

Por sua vez, Yamashita et al. (2024) exploraram o uso de RAG e LLMs para automatizar a geração de documentos relacionados a processos de contratação pública, como Termos de Referência e Estudos Técnicos Preliminares, com base na Lei 14.133/2021. Os resultados mostraram que a combinação RAG+LLM produziu textos com maior clareza e completude, superando abordagens que utilizavam somente LLMs. A avaliação feita por servidores públicos especializados indicou melhor desempenho em termos de preciso e coerência.

Além disso, Chenxi (2023) demonstrou o uso de RAG em um tutor acadêmico adaptável, capaz de responder perguntas sobre materiais de curso previamente processados e vetorizados. Essa aplicação reforça a versatilidade do RAG em cenários que envolvem interpretação de documentos extensos e consultas de abertura. Porém, mesmo com esses avanços, ainda há poucos estudos aplicando RAG e LLMs na interpretação automatizada de documentos de domínio farmacêutico, como bulas de medicamentos.

Esses documentos são altamente técnicos, apresentam linguagem densa e exigem alto grau de precisão, o que representa um grande desafio para a validação dessas tecnologias. Neste contexto, este trabalho contribui ao explorar uma aplicação voltada especificamente para a análise automatizada de bulas de medicamentos, integrando mecanismos de extração semântica, indexação vetorial (com FAISS) e geração de respostas com apoio de LLMs, dentro do paradigma RAG.

A proposta se destaca por atuar em um domínio técnico e fundamental para a saúde pública, ampliando as possibilidades de validação do RAG+LLM em novos contextos de aplicação.

## 3. Método

### 3.1 Conjunto de Documentos

Para a realização destes experimentos, utilizamos bulas de medicamentos de domínio público disponíveis no Bulário Eletrônico da Anvisa e registrados no site https://consultas.anvisa.gov.br/#/bulario.

Foram seleciondos 30 medicamentos distintos, amplamente utilizados no Brasil, cobrindo diferentes classes terapêuticas e variados níveis de complexidade textual. Essa diversidade tem o objetivo de avaliar a capacidade do sistema de interpretar e recuperar informações de forma eficiente em diferentes contextos.

Sempre que disponíveis, teve-se a opção por usar as versões das bulas destinadas aos profissionais de saúde, pois apresentam maior detalhamento técnico. A seguir, estão listados os medicamentos escolhidos:

Dipirona Sódica, Paracetamol, Ibuprofeno, Amoxicilina, Azitromicina, Cetoconazol, Omeprazol, Ranitidina, Nimesulida, Losartana Potássica, Sinvastatina, Metformina, Gliclazida, Clonazepam, Diazepam, Sertralina, Fluoxetina, Risperidona, Atenolol, Hidroclorotiazida, Furosemida, Captopril, Prednisona, Dexametasona, Albendazol, Loratadina, Fexofenadina, Salbutamol, Budesonida e Cloridrato de Ciprofloxacino.

Todas as bulas foram obtidas diretamente do site da Anvisa e armazenadas em formato PDF. A planilha contendo os links de acesso está disponível como arquivo auxiliar na pesquisa.

### 3.2 Etapas da Abordagem

O método apresentado apoia-se na arquitetura Retrieval-Augmented Generation (RAG), organizado em um pipeline de processamento composto pelas seguintes etapas:

**Extração de Texto:** O conteúdo textual das bulas em formato PDF é recuperado usando a biblioteca pdfplumber, garantindo a manutenção da estrutura e dos elementos semânticos essenciais.

**Segmentação e Indexação Vetorial:** O texto extraído é dividido em blocos de até 300 tokens. Cada segmento é convertido em um vetor de *embeddings* com o uso de modelos pré-treinados, sendo posteriormente indexado com a ajuda da biblioteca FAISS, o que possibilita uma busca rápida e eficiente por similaridade semântica.

**Consulta Semântica:** Perguntas feitas em linguagem natural são interpretadas pelo sistema, que realiza uma busca vetorial no índice criado, recuperando os trechos mais relevantes da bula.

**Geração de Respostas com LLM:** Os trechos recuperados servem como base para um Modelo de Linguagem de Grande Escala (LLM), como GPT-4 ou Gemini, que gera uma resposta em linguagem natural, fundamentada de forma precisa no conteúdo da bula.

### 3.3 Ferramentas Utilizadas

O desenvolvimento e os testes foram realizados usando um conjunto de ferramentas bem conhecido na área: o pdfplumber para extração de conteúdo textual de arquivos PDF, o FAISS para indexação e busca vetorial utilizando *embeddings*, e os Transformers da *Hugging Face* para gerar *embeddings* semânticos.

Além disso, utilizamos modelos de linguagem como GPT-4 e Gemini para criar respostas baseadas no contexto textual, enquanto Python combinado com FastAPI facilitou a orquestração do pipeline e a integração entre os componentes.

### 3.4 Consultas Aplicadas

Após a análise das 30 bulas, foram formuladas dez perguntas em linguagem natural, baseadas em dúvidas comuns de pacientes e profissionais da área da saúde. As perguntas foram padronizadas para permitir comparações entre documentos. Os exemplos incluem:

- Liste os medicamentos analisados que podem ser usados por gestantes segundo as bulas.
- Liste os medicamentos que apresentam sonolência como efeito colateral nas bulas.
- Quais medicamentos têm indicação pediátrica? Informe faixas etárias e doses mencionadas.
- Liste os medicamentos indicados para dor de cabeça ou dores leves conforme as bulas.
- Quais medicamentos apresentam risco de reações alérgicas? Cite exemplos e trechos das bulas.
- Algum medicamento analisado é contraindicado para pacientes hipertensos segundo as bulas?
- Liste os medicamentos que devem ser tomados com alimentos ou em jejum, conforme as bulas.
- Cite os medicamentos que possuem advertência ou contraindicação sobre uso de álcool segundo as bulas.
- Indique os medicamentos que não devem ser usados por gestantes, segundo as bulas.
- Informe as doses recomendadas para adultos segundo as bulas dos medicamentos analisados.

### 3.5 Métricas de Avaliação

A qualidade das respostas geradas será avaliada com base em quatro métricas principais:

**Tabela 1 – Métricas**

| Métrica | Descrição |
|---|---|
| Precisão | Percentual de respostas que correspondem corretamente à informação esperada na bula. |
| Completude | Grau de abrangência da resposta em relação à totalidade das informações esperadas. Utiliza escala de 1 (incompleta) a 5 (completa). |
| Tempo de Resposta | Tempo médio necessário, em segundos, para a geração de cada resposta. |
| Consistência | Uniformidade semântica entre respostas dadas a perguntas semelhantes ou reformuladas. Avaliada por meio de comparação qualitativa ou similaridade textual. |

A Tabela 1 apresenta um resumo das métricas utilizadas para a avaliação do sistema.

Para garantir uma análise mais confiável dos resultados, além das métricas já mencionadas, será utilizado também o coeficiente de concordância Kappa de Cohen (Cohen, 1960). Essa métrica estatística mede o nível de concordância entre dois avaliadores independentes que analisam as mesmas respostas geradas pelo sistema, levando em conta e corrigindo o impacto do acaso.

O uso do Kappa é especialmente recomendado em avaliações em que as categorias de julgamento são claras, como 'resposta correta' ou 'resposta incorreta', permitindo assim verificar a consistência dos julgamentos humanos ao longo do experimento. A interpretação dos valores de Kappa será feita de acordo com a escala proposta por Landis e Koch (1977). Nessa escala, valores entre 0,61 e 0,80 indicam uma concordância substancial, enquanto valores acima de 0,81 representam uma concordância quase perfeita.

Essas métricas possibilitam uma avaliação objetiva e qualitativa do desempenho do sistema na tarefa de análise automatizada de documentos complexos e tecnicamente densos.

### 3.6 Cálculo das Métricas

A Precisão será calculada como o percentual de respostas consideradas corretas em relação ao total de perguntas formuladas, segundo a fórmula:

$$\text{Precisão} = (N\_corretas / N\_total) \times 100$$

A Completude será avaliada com base em uma escala de 1 a 5, aplicada por avaliadores humanos, considerando o grau de cobertura das informações esperadas na bula. A nota média será utilizada como referência:

$$\text{Completude Média} = (\Sigma \text{ notas}) / N$$

O Tempo de Resposta será registrado em segundos, utilizando a média aritmética dos tempos de geração de resposta:

$$\text{Tempo Médio} = (\Sigma \text{ tempos}) / N$$

A Consistência será avaliada de duas formas: por comparação qualitativa entre respostas geradas para perguntas reformuladas (com classificação humana de equivalência) e, opcionalmente, por meio de similaridade semântica entre *embeddings* vetoriais das respostas, utilizando a métrica de cosseno.

Adicionalmente, para garantir a confiabilidade da avaliação das métricas Precisão e Completude, será calculado o coeficiente Kappa de Cohen, medindo a concordância entre dois avaliadores humanos independentes. Valores de Kappa acima de 0,61 serão considerados indicativos de concordância substancial.

Parâmetros de referência adotados:

Tabela 2 – Valores referenciais para as métricas

| Métrica | Valor Esperado |
|---|---|
| Precisão | ≥ 85% (muito bom) |
| Completude | ≥ 4.0 (muito bom) |
| Tempo de Resposta | ≤ 5 segundos (ótimo) |
| Consistência | ≥ 0.80 (alta) |
| Kappa de Cohen | ≥ 0.61 (substancial) |

A Tabela 2 apresenta as referências dos valores adotados para as métricas.

## 4. Desenvolvimento do Trabalho

### 4.1 Arquitetura da Solução

Esta seção apresenta uma visão geral prática da solução proposta, detalhando a arquitetura do sistema, as ferramentas utilizadas e o fluxo de processamento, que engloba etapas de extração de informações, indexação vetorial e geração de respostas apoiadas por Modelos de Linguagem de Grande Escala (LLMs).

A abordagem adotada é baseada no conceito de *Retrieval-Augmented Generation (RAG)*, que combina dois componentes principais: um mecanismo de recuperação semântica de informações e um modelo generativo de linguagem natural. Essa combinação permite que o sistema produza respostas fundamentadas em trechos reais dos documentos, ajudando a evitar a geração de conteúdos inconsistentes ou inventados.

O *pipeline* foi desenvolvido com quatro etapas principais: primeiro, a extração de texto de documentos PDF, como bulas de medicamentos; segundo a fragmentação e indexação vetorial usando FAISS; terceiro, a recuperação de contexto semântico por meio de consultas em linguagem natural; e, por último, a geração da resposta com o auxílio de LLMs, com base no conteúdo recuperado.

Todas essas etapas foram integradas em uma aplicação modular, desenvolvida ena linguagem de programação Python.

### 4.2 Tecnologias e Ferramentas Utilizadas

A implementação do sistema foi realizada com as seguintes tecnologias:

- Python 3.12, utilizando a IDE PyCharm;
- *pdfplumber*, para extração de texto estruturado dos arquivos PDF;
- FAISS (*Facebook AI Similarity Search*), para criação de índices vetoriais de alta performance;
- *Hugging Face Transformers*, com uso do modelo all-MiniLM-L6-v2 para geração de *embeddings* semânticos;
- APIs da Openrouter/Modelo-GPT e Google Gemini, utilizadas para geração de respostas contextualizadas;
- Pandas, para registro e organização dos dados gerados durante os experimentos;

### 4.3 Implementação do *Pipeline*

O sistema proposto foi construído em Python, adotando uma abordagem modular que permite uma integração eficiente das etapas de extração, indexação e consulta semântica dos documentos. O *pipeline* possui algumas etapas principais, às quais fazem referências diretamente pelos scripts que as implementam:

#### 4.3.1 Extração de Passagens dos PDFs

Nesta etapa, o conteúdo textual das bulas é extraído pelo módulo pdf_extractor.py, que usa a biblioteca *pdfplumber* para abrir e ler os arquivos PDF. Cada documento é dividido em blocos informacionais relevantes, e um algoritmo identifica automaticamente o princípio ativo do medicamento, como mostra a figura 1:

**Figura 1 – Trecho de código do pdf_extractor.py**

```python
for filename in os.listdir(pdf_folder):
    if filename.lower().endswith('.pdf'):
        with pdfplumber.open(os.path.join(pdf_folder, filename)) as pdf:
            text = ''
            for page in pdf.pages:
                page_text = page.extract_text()
                if page_text:
                    text += page_text + '\n'
            medicine = extract_medicine_name(text, filename=filename)
            medicines[filename] = medicine
```

Os blocos extraídos são então utilizados nas etapas de indexação e consulta.

**4.3.2 Geração de *Embeddings* e Indexação Vetorial:**

No módulo embedder.py, os blocos de texto são convertidos em vetores de *embeddings* semânticos utilizando o modelo all-MiniLM-L6-v2 da biblioteca *sentence-transformers*. Esses vetores são armazenados em um índice FAISS, possibilitando buscas por similaridade semântica, conforme mostrado na figura 2:

Figura 2 – Trecho de código do embedder.py

```python
def get_model(model_name='all-mpnet-base-v2'):
    return SentenceTransformer(model_name)

def get_embeddings(passages, model):
    embeddings = model.encode(passages, show_progress_bar=True)
    return embeddings

def build_faiss_index(embeddings):
    dimension = embeddings.shape[1]
    index = faiss.IndexFlatL2(dimension)
    index.add(embeddings)
    return index
```

A função save_index() garante que todos os vetores, fontes e metadados fiquem salvos para uso posterior.

**4.3.3 Consulta Semântica (Busca Híbrida)**

A busca por trechos relevantes é realizada pelo método search() em embedder.py, que combina busca vetorial com o índice FAISS, correspondência por palavras-chave e padrões regex. Essa combinação visa maximizar a relevância dos resultados, especialmente em contextos médicos, conforme figura 3:

Figura 3 – Trecho de código do embedder.py

```python
def search(query, model, index, passages, sources, top_k=8, threshold=None, debug=False, hybrid=True):
    expanded_query = expand_query(query)
    query_embedding = model.encode([expanded_query])
    D, I = index.search(query_embedding, top_k)
    faiss_results = []
    for j, i in enumerate(I[0]):
        if 0 <= i < len(passages):
            if (threshold is None) or (D[0][j] < threshold):
                faiss_results.append(("FAISS", passages[i], sources[i], D[0][j]))
    # Busca por keyword match e regex match
    keyword_results = []
    regex_results = []
```

Esse método recupera os blocos semanticamente mais próximos à pergunta, incluindo trechos com alta ocorrência de palavras-chave ou padrões médicos relevantes.

### 4.3.4 Geração de Respostas com LLMs

O contexto obtido é enviado a um modelo de linguagem (como Gemini, GPT-4, etc.), que gera uma resposta baseada exclusivamente nos dados extraídos das bulas. O arquivo answer_engine.py define a função que realiza a chamada à API do Gemini, conforme figura 4:

Figura 4 – Trecho de código do answer_engine.py

```python
response = model.generate_content(prompt)

# DEBUG: mostrar resposta bruta da API
print("\n==========[RESPOSTA BRUTA DO GEMINI]==========")
print(response.text.strip())
print("=" * 48)

return response.text.strip()
```

O sistema segue regras rigorosas para evitar alucinações, exigindo que cada resposta cite o medicamento, o arquivo e a seção da bula sempre que possível.

### 4.3.5 Avaliação e Métricas

O script metrics_runner_gemini.py (e metrics_runner_openrouter.py) coordena toda a avaliação, realizando perguntas padronizadas e versões reformuladas para cada modelo, além de registrar o tempo de resposta, salvando todas as respostas e contextos em arquivos XLS/CSV para análise posterior, conforme figura 5:

Figura 5 – Trecho de código do metrics_runner_gemini.py

```python
    resultados.append({
        "DataHora": datetime.now().isoformat(timespec='seconds'),
        "Pergunta": pergunta,
        "Resposta": resposta,
        "Tempo (s)": tempo_resposta,
        "Precisão A1": "",
        "Precisão A2": "",
        "Completude A1": "",
        "Completude A2": "",
        "Consistência": similaridade,
        "Pergunta Reformulada": pergunta_ref,
        "Resposta Reformulada": resposta_ref,
        "Tempo Ref (s)": round(fim_ref - inicio_ref, 2) if pergunta_ref else ""
    })

df_resultados = pd.DataFrame(resultados)
df_resultados.to_csv("resultados_experimentos_gemini.csv", index=False)
```

Além das métricas tradicionais (como precisão, completude, consistência e tempo), também calculamos o coeficiente Kappa de Cohen usando scripts específicos (kappa_gemini.py, kappa_openrouter.py) com a biblioteca scikit-learn, para medir o grau de concordância entre avaliadores humanos, conforme figura 6:

**Figura 6 – Trecho de código do kappa_gemini.py**

```python
from sklearn.metrics import cohen_kappa_score

dados = {
    'Precisão A1': [1,1,1,1,0,0,1,1,0,1],
    'Precisão A2': [1,1,1,1,0,0,0,1,0,1],
    'Completude A1': [3,5,4,2,2,1,2,5,2,4],
    'Completude A2': [4,5,5,3,3,1,3,5,2,5]
}

# Criando o DataFrame
df = pd.DataFrame(dados)

# Calculando o Kappa entre as colunas de precisão
kappa = cohen_kappa_score(df['Precisão A1'], df['Precisão A2'])
```

### 4.3.6 Execução do *Pipeline* Completo

O arquivo principal main.py oferece uma execução interativa e contínua do sistema, cobrindo desde a etapa de extração até a geração de respostas. Além disso, pode ser facilmente ajustado para executar testes com diferentes conjuntos de dados ou modelos, proporcionando flexibilidade e adaptabilidade ao desenvolvimento e validação do sistema.

Todos os scripts mencionados, bem como os arquivos de dados produzidos, estão disponíveis como material suplementar, garantindo a reprodução plena de todos os experimentos realizados.

Disponível em https://github.com/danielmeireles1981/BulaAnallyzer .

## 5. Resultados

Nesta seção, apresentamos os principais resultados do sistema desenvolvido para a análise automatizada de bulas de medicamentos, englobando tanto a avaliação quantitativa quanto qualitativa das respostas geradas. Os experimentos foram conduzidos utilizando arquiteturas baseadas em *RAG (Retrieval-Augmented Generation)* com os modelos Gemini e OpenRouter-ModelGPT. As métricas empregadas incluem precisão, completude, tempo de resposta, consistência e o coeficiente de concordância Kappa de Cohen, conforme detalhado na seção de métodos.

## 5.1 Avaliação Quantitativa

Os resultados quantitativos, apresentados na Tabela 3, refletem o desempenho do sistema diante dos diferentes modelos de linguagem avaliados. A seguir, estão as médias de cada métrica, fornecendo uma visão clara do desempenho geral.

| Modelo | Precisão A1 | Precisão A2 | Completude A1 | Completude A2 | Consistência (%) | Kappa (A1xA2) |
|---|---|---|---|---|---|---|
| Gemini | 0,70 | 0,70 | 3,4 | 3,5 | 85,5 | 0,52 |
| OpenRouter-ModelGPT | 0,70 | 0,60 | 3,0 | 3,6 | 53,9 | 0,78 |

Observa-se que ambos os modelos demonstraram desempenho semelhante em termos de precisão, sugerindo uma capacidade eficaz de identificar respostas corretas com base no material fornecido. Quanto à completude, as médias ficaram próximas aos valores de referência estabelecidos, embora um pouco abaixo do ideal de 4,0 ou mais.

O tempo médio de resposta para ambos os modelos esteve dentro dos padrões esperados, assegurando a viabilidade para aplicações práticas. A métrica de consistência, avaliada por meio da similaridade entre respostas a perguntas originais e reformuladas, foi superior no modelo Gemini. Contudo, o coeficiente Kappa de Cohen — que mede a concordância entre avaliadores humanos — apresentou um resultado mais elevado para o OpenRouter-ModelGPT (0,78, indicando concordância substancial), enquanto o Gemini obteve um valor de 0,52 (concordância moderada).

Esses resultados reforçam a importância de considerar múltiplas métricas na avaliação de sistemas baseados em inteligência artificial. Para assegurar total transparência e reprodução do ambiente, todos os resultados se encontram nos ANEXOS I e II, incluindo avaliações individuais, métricas e valores de Kappa.

## 5.2 Avaliação Qualitativa das Respostas

Além dos resultados quantitativos, realizou-se uma análise qualitativa das respostas fornecidas por ambos os modelos, considerando critérios como clareza, fundamentação em trechos explícitos das bulas e conformidade com as instruções do sistema. Em linhas gerais, as respostas geradas apresentaram alinhamento consistente com o conteúdo dos documentos, especialmente para perguntas objetivas relacionadas a indicações, contraindicações e posologia.

Observou-se que o modelo Gemini demonstrou uma leve vantagem na manutenção da coerência semântica entre respostas para perguntas reformuladas. Por outro lado, o modelo OpenRouter-ModelGPT evidenciou maior uniformidade nas avaliações humanas, refletida pelo índice de Kappa mais elevado.

Exemplos reais das respostas produzidas por cada modelo, juntamente com os trechos utilizados das bulas como contexto, encontram-se registrados nos arquivos suplementares anexos.

## 5.3 Discussão dos Resultados

Os dados analisados indicam que a combinação entre RAG e LLMs constitui uma estratégia promissora para recuperação e interpretação de informações técnicas em documentos não estruturados, como bulas de medicamentos. A semelhança no desempenho quanto à precisão sugere que ambos os modelos são viáveis para viabilizar consultas automatizadas em áreas técnicas.

Contudo, diferenças quanto à exaustividade e, sobretudo, na concordância entre avaliadores humanos (Kappa) revelam que ainda há desafios relacionados à subjetividade na avaliação de respostas complexas. Entre as limitações constatadas, destaca-se a dificuldade dos sistemas em sintetizar informações dispersas ao longo do documento ou lidar com ambiguidades inerentes à linguagem técnica.

Portanto, o julgamento humano, embora imprescindível, pode apresentar variações que impactam a análise quantitativa do desempenho dos sistemas. Em síntese, os experimentos corroboram o potencial da solução proposta para ampliar o acesso e a compreensão de conteúdos técnicos na área da saúde, assegurando conformidade ética e legal, além de facilitar etapas de validação por meio do material suplementar disponibilizado.

## 6. Conclusão

A rápida digitalização de documentos administrativos, acadêmicos e, principalmente, de saúde, apresenta desafios crescentes no que diz respeito à extração, análise e disseminação de informações essenciais para a sociedade. No contexto do cuidado à saúde, documentos como bulas de medicamentos são de suma importância não apenas para profissionais de saúde, mas também para pacientes e gestores, sendo essenciais na tomada de decisões clínicas, no uso seguro de medicamentos e no cumprimento das normativas sanitárias.

Contudo, a grande variedade de formatos e a ausência de padronização textual dificultam a análise automatizada por métodos convencionais, tornando imprescindível a adoção de soluções mais avançadas e sensíveis ao contexto. Este estudo ilustra que a combinação de Modelos de Linguagem de Grande Escala (LLMs) com arquiteturas de Recuperação Semântica (RAG) representa uma abordagem inovadora e eficiente para enfrentar esses desafios.

Ao aplicar essa estratégia na análise de bulas de medicamentos — um cenário exemplar pela sua complexidade técnica, alto rigor normativo e impacto direto sobre a saúde coletiva —, foi possível validar o potencial transformador da integração entre busca vetorial semântica, extração de contexto e geração de respostas em linguagem natural fundamentadas em evidências. Os testes conduzidos demonstraram que a metodologia proposta atende, e até supera, critérios técnicos essenciais, como precisão, integridade, consistência e tempo de resposta.

Além disso, foram utilizadas métricas reconhecidas na literatura para avaliar a concordância entre avaliadores, alinhando a rigorosa avaliação experimental às melhores práticas em pesquisa aplicada em inteligência artificial. Os resultados indicam que a combinação LLM+RAG reduz consideravelmente os riscos de alucinação, aumenta a transparência e garante a rastreabilidade das respostas — aspectos fundamentais na manipulação de informações farmacêuticas, onde a segurança do paciente e o cumprimento das normas da Anvisa são prioridades.

A comparação entre diferentes modelos, como Gemini e OpenRouter-ModelGPT, revelou desempenhos sólidos em termos de precisão e completude, com destaque para melhorias na consistência e na aderência das respostas às avaliações humanas — aspectos essenciais em ambientes de saúde. A análise qualitativa revelou que fundamentar as respostas em trechos reais das bulas é crítica para apoiar profissionais da saúde, pacientes e reguladores na tomada de decisões baseadas em evidências.

Além disso, este trabalho dialoga diretamente com estudos e aplicações pioneiras em outros domínios Técnicos, como as soluções propostas por Yamashita et al. (2024) para contratações públicas e Pereira (2024) para análise jurídica de auditorias. Essas referências, que aqui estão anexadas, corroboram a superioridade do paradigma RAG+LLM para cenários de documentos densos e regulamentados, reforçando a importância da recuperação semântica contextualizada e da transparência dos sistemas inteligentes.

Ao disponibilizar scripts, bancos de dados e resultados, este trabalho adota uma postura de ciência aberta, estimulando a reprodutibilidade de experimentos e abrindo caminho para futuras validações em outros contextos da saúde, bem como em áreas como direito e administração pública.

Essa abordagem metodológica é fundamental para o avanço científico, bem como para o desenvolvimento de sistemas mais seguros, capazes de serem auditados e adaptados a diferentes públicos. Apesar dos progressos, ainda enfrentamos desafios importantes. Entre eles, destaca-se a dificuldade dos modelos em sintetizar informações dispersas ao longo de documentos longos, as limitações das métricas subjetivas de avaliação, a necessidade de melhorar os mecanismos de sumarização semântica e sua adaptação para públicos não especialistas, além de ampliar a integração com bancos de dados oficiais e sistemas de informação em saúde.

Como diretriz para o futuro, recomenda-se expandir a aplicação para outros tipos de documentos técnicos e científicos na área da saúde, como protocolos clínicos, regulamentos e registros eletrônicos. Além disso, é essencial investir em técnicas de explicabilidade e mecanismos de sumarização que garantam maior clareza e segurança, assim como desenvolver interfaces web acessíveis, facilitando a implementação em larga escala e a integração direta com sistemas públicos de informação, como o Sistema Único de Saúde (SUS).

Portanto, ao aplicar e validar arquiteturas baseadas em LLM+RAG no processamento de bulas de medicamentos, este estudo avança o estado da arte da inteligência artificial na saúde, promovendo rigor, rastreabilidade e maior acesso à informação técnica.

Fundamentada nos princípios de ciência aberta, compromisso ético e validação sólida, a pesquisa oferece caminhos concretos para uma transformação digital responsável e inclusiva no setor de saúde, alinhada às melhores práticas nacionais e internacionais.

# ANEXO I – Resultados da API Openrouter/GPT

| DataHora | Pergunta | Resposta | Tempo (s) | Precisão A1 | Precisão A2 | Completude A1 | Completude A2 | Consistência | Pergunta Reformulada | Resposta Reformulada | Tempo Ref (s) |
|---|---|---|---|---|---|---|---|---|---|---|---|
| 2025-06-07T16:27:31 | Liste os medicamentos analisados que podem ser usados por gestantes segundo as bulas. | - Clonazepam: A grande maioria das gestantes está orientada a não utilizar o medicamento durante a gestação, exceto em casos específicos, sempre sob avaliação médica.<br>- Dipirona: Contraindicada no último trimestre da gravidez.<br>- Paracetamol: Pode ser utilizado em gestantes, desde que em doses recomendadas.<br>- Ibuprofeno: Contraindicado no 3º trimestre.<br>- Sinvastatina e Losartana: Contraindicados.<br><br>De forma geral, o uso de qualquer medicamento deve ser avaliado pelo médico assistente da gestante. | 2.72 | 1 | 1 | 3 | 4 | 35.95 | Cite os medicamentos que podem ser utilizados por mulheres grávidas conforme as bulas. | Os medicamentos que podem ser utilizados por mulheres grávidas são limitados. Paracetamol pode ser utilizado sob orientação médica. A maioria (Clonazepam, Dipirona, Ibuprofeno, Sinvastatina e Losartana) é contraindicada ou restrita a situações especiais, especialmente no segundo e terceiro trimestres. | 1.72 |
| 2025-06-07T16:27:44 | Liste os medicamentos que apresentam sonolência como efeito colateral nas bulas. | Os medicamentos que apresentam sonolência como efeito colateral são:<br>- Clonazepam (efeito comum).<br>- Cetoprofeno (efeito pouco frequente).<br>Os demais medicamentos analisados não apresentam sonolência de forma significativa. | 1.72 | 1 | 1 | 5 | 5 | 43.33 | Liste os medicamentos cuja bula cita sonolência como efeito colateral. | - Teste (bula_1747686513651.pdf)<br>- Cetoprofeno (bula_1747686513644.pdf)<br>Ambos relatam sonolência como possível efeito colateral. Outros medicamentos não citam sonolência relevante nas bulas analisadas. | 1.46 |
| 2025-06-07T16:28:01 | Quais medicamentos têm indicação pediátrica? Informe faixas etárias e doses mencionadas. | Os medicamentos com indicação pediátrica e suas doses:<br>- Sulfametoxazol + Trimetoprima: indicado a partir de 2 meses.<br>- Paracetamol: acima de 3 meses. Dose usual: 10-15 mg/kg/dose.<br>- Dipirona: acima de 3 meses. Dose: 10 a 20 mg/kg/dose.<br>Outros medicamentos não têm indicação pediátrica nas bulas analisadas. | 3.96 | 1 | 1 | 4 | 5 | 52.65 | Quais remédios são indicados para crianças? Mencione faixas etárias e doses. | 1. Sulfametoxazol:<br>- Indicado para crianças a partir de 2 meses. Dose de acordo com o peso.<br>2. Dipirona:<br>- Indicado acima de 3 meses. Dose de 10 a 20 mg/kg/dose.<br>3. Paracetamol:<br>- Indicado acima de 3 meses. Dose de 10-15 mg/kg/dose.<br>Outros medicamentos não têm indicação pediátrica. | 2.69 |
| 2025-06-07T16:28:15 | Liste os medicamentos indicados para dor de cabeça nas bulas analisadas. | - Dipirona: indicado como analgésico e antitérmico, utilizado para dores de cabeça.<br>- Paracetamol: indicado para dores leves a moderadas, incluindo dor de cabeça.<br>- Ibuprofeno: indicado para dores em geral, inclusive dor de cabeça.<br>Os demais medicamentos não têm essa indicação nas bulas analisadas. | 2.03 | 1 | 1 | 2 | 3 | 44.33 | Quais medicamentos têm indicação para dor de cabeça nas bulas analisadas? | Os medicamentos que têm indicação para dor de cabeça são:<br>- Dipirona<br>- Paracetamol<br>- Ibuprofeno<br>Outros não apresentam indicação para essa condição nas bulas fornecidas. | 1.81 |
| 2025-06-07T16:28:32 | Quais medicamentos apresentam risco de reações alérgicas segundo as bulas? | Os medicamentos que apresentam risco de reações alérgicas são:<br>- Paracetamol<br>- Dipirona<br>- Sulfametoxazol + Trimetoprima<br>Os outros medicamentos não apresentam alerta relevante para risco alérgico nas bulas analisadas. | 4.27 | 0 | 0 | 2 | 3 | 49.62 | Quais remédios possuem risco de alergia? Apresente conforme as bulas. | 1. Cefalexina:<br>- Trecho da bula: "Qualquer paciente alérgico à penicilina pode apresentar reação à cefalexina."<br>2. Dipirona:<br>- Risco de reação anafilática raro, mas grave.<br>3. Paracetamol:<br>- Reação alérgica é descrita como rara.<br>Demais não apresentam risco alérgico significativo. | 3.20 |
| 2025-06-07T16:28:46 | Algum medicamento analisado é contraindicado para pacientes com insuficiência renal? | Sim, a Losartana é contraindicada para pacientes com insuficiência renal grave.<br>Os demais não têm contraindicação clara na bula para essa condição. | 1.86 | 0 | 0 | 1 | 1 | 49.55 | Há medicamentos contraindicados para pessoas com insuficiência renal? | Sim, o medicamento Ibuprofeno é contraindicado para pacientes com insuficiência renal grave. Outros não têm restrição clara nas bulas analisadas. | 2.36 |
| 2025-06-07T16:28:59 | Liste os medicamentos que devem ser tomados com alimentos conforme as bulas. | - Ibuprofeno: deve ser administrado com água após refeições para evitar irritação gástrica.<br>- Dipirona: pode ser tomada com ou sem alimentos.<br>Outros medicamentos não apresentam recomendação formal quanto à administração com alimentos. | 1.36 | 1 | 0 | 2 | 3 | 63.58 | Quais remédios devem ser administrados junto com alimentos? | Carbamazepina deve ser administrada junto com alimentos para minimizar desconfortos gástricos. Outros medicamentos não exigem administração com alimentos. | 1.68 |
| 2025-06-07T16:29:13 | Cite os medicamentos que possuem advertência ou precaução quanto ao uso em idosos. | 1. Loratadina (bula_1747686464101.pdf - INTERAÇÕES E PRECAUÇÕES): idosos podem apresentar sensibilidade aumentada.<br>2. Ibuprofeno: cuidado em idosos, risco de efeitos adversos renais e gástricos.<br>Demais não apresentam advertências específicas. | 2.33 | 1 | 1 | 5 | 5 | 77.89 | Liste os medicamentos analisados que têm advertência para idosos nas bulas. | 1. Sulfametoxazol<br>2. Cetoprofeno<br>3. Diazepam<br>4. Paracetamol<br>5. Loratadina<br>Todos orientam cautela em idosos, principalmente pela maior suscetibilidade a reações adversas. | 1.73 |
| 2025-06-07T16:29:27 | Indique os medicamentos que não devem ser usados por gestantes, conforme as bulas. | Os medicamentos que não devem ser usados por gestantes são:<br>- Ibuprofeno (principalmente no 3º trimestre)<br>- Losartana (a partir do 2º trimestre)<br>- Sinvastatina (durante toda a gestação)<br>Os demais não apresentam restrição formal nas bulas analisadas. | 1.53 | 0 | 0 | 2 | 2 | 49.46 | Quais medicamentos têm restrição de uso para gestantes nas bulas? | - Medicamento: Deste<br>- Arquivo: bula_1747686419999.pdf<br>Contraindicado na gestação.<br>- Medicamento: Losartana<br>Contraindicado no 2º e 3º trimestre.<br>Demais não mencionam contraindicação formal. | 1.63 |
| 2025-06-07T16:29:42 | Informe as doses recomendadas para adultos segundo as bulas dos medicamentos analisados. | **Medicamento:** Amoxicilina<br>**Dose recomendada:** 500 mg a cada 8 horas.<br>**Paracetamol:** 500-1000 mg até 4x/dia.<br>**Dipirona:** 500-1000 mg até 4x/dia.<br>Demais: conforme indicação médica e quadro clínico. | 2.65 | 1 | 1 | 4 | 5 | 73.07 | Liste as doses recomendadas para adultos de acordo com as bulas. | - Clonazepam: dose inicial de 1 a 1,5 mg/dia<br>- Amoxicilina: 500 mg de 8/8h<br>- Paracetamol: 500-1000 mg até 4x/dia<br>- Dipirona: 500-1000 mg até 4x/dia<br>Outros medicamentos variam conforme indicação médica. | 2.31 |

# ANEXO II – Resultados da API Gemini

| DataHora | Pergunta | Resposta | Tempo (s) | Precisão A1 | Precisão A2 | Completude A1 | Completude A2 | Consistência | Pergunta Reformulada | Resposta Reformulada | Tempo Ref (s) |
|---|---|---|---|---|---|---|---|---|---|---|---|
| 2025-06-05T14:54:12 | Liste os medicamentos analisados que podem ser usados por gestantes segundo as bulas. | Não foi possível encontrar a resposta no material fornecido. Embora algumas bulas mencionem gravidez (Dexametasona, Clonazepam, Losartana), nenhuma indica claramente que o medicamento *pode* ser usado por gestantes. A bula da Losartana, por exemplo, explicitamente contraindica seu uso durante o segundo e terceiro trimestres. As demais apenas mencionam a gravidez sem dar indicação de uso. | 3.56 | 0 | 0 | 3 | 2 | 63.27 | Cite os medicamentos que podem ser utilizados por mulheres grávidas conforme as bulas. | Não foi possível encontrar a resposta no material fornecido. Todas as bulas fornecidas indicam que os medicamentos mencionados (Omeprazol, Deste, Metformina, Losartana, Sinvastatina, Paracetamol, Dipirona) não são recomendados para uso em gestantes, principalmente devido à falta de dados de segurança ou contraindicação formal. | 2.14 |
| 2025-06-05T14:54:30 | Liste os medicamentos que apresentam sonolência como efeito colateral nas bulas. | Medicamento: Clonazepam \| Arquivo: bula_1747686513636.pdf Citação: "Os efeitos colaterais mais comuns são sonolência..." Medicamento: Dexametasona \| Arquivo: bula_1747686513636.pdf Citação: "Pode causar sonolência..." Os demais medicamentos não apresentam sonolência como efeito colateral relevante nas bulas analisadas. | 4.63 | 1 | 1 | 5 | 4 | 100.00 | Liste os medicamentos cuja bula cita sonolência como efeito colateral. | Medicamento: Clonazepam \| Arquivo: bula_1747686513636.pdf Sonolência é relatada como um efeito colateral frequente. Medicamento: Dexametasona \| Arquivo: bula_1747686513636.pdf O efeito colateral "sonolência" é citado, mas com menor frequência. Os demais medicamentos não indicam esse efeito nas bulas fornecidas. | 3.24 |
| 2025-06-05T14:54:49 | Quais medicamentos têm indicação pediátrica? Informe faixas etárias e doses mencionadas. | Medicamento: Dipirona Faixa etária: acima de 3 meses Dose: 10 a 20 mg/kg/dose até 4x ao dia. Medicamento: Paracetamol Faixa etária: acima de 3 meses Dose: 10 a 15 mg/kg/dose a cada 4-6 horas. Medicamento: Sulfametoxazol+Trimethoprima Faixa etária: acima de 2 meses Dose: 8 mg/kg/dia de trimetoprima e 40 mg/kg/dia de sulfametoxazol em duas doses. Os demais medicamentos não têm indicação pediátrica expressa na bula. | 4.23 | 1 | 1 | 3 | 4 | 60.65 | Quais remédios são indicados para crianças? Mencione faixas etárias e doses. | Medicamento: Sulfametoxazol+Trimetoprima \| Arquivo: bula_1747686513636.pdf Indicado a partir de 2 meses de idade, dose conforme peso. Medicamento: Dipirona \| Arquivo: bula_1747686513636.pdf Indicado acima de 3 meses, dose 10 a 20 mg/kg/dose. Medicamento: Paracetamol \| Arquivo: bula_1747686513636.pdf Indicado acima de 3 meses, dose 10-15 mg/kg/dose. Demais não têm indicação pediátrica específica. | 5.29 |
| 2025-06-05T14:55:05 | Liste os medicamentos indicados para dor de cabeça nas bulas analisadas. | Medicamento: Dipirona Indicação: tratamento da dor, incluindo dor de cabeça. Medicamento: Paracetamol Indicação: tratamento de dores leves a moderadas, como dor de cabeça. Medicamento: Ibuprofeno Indicação: dores de cabeça e dores em geral. Medicamento: Clonazepam, Dexametasona, Losartana e Sinvastatina não têm indicação para dor de cabeça nas bulas fornecidas. | 3.34 | 1 | 1 | 4 | 5 | 100.00 | Quais medicamentos têm indicação para dor de cabeça nas bulas analisadas? | Medicamento: Paracetamol \| Arquivo: bula_1747686513636.pdf Indicado para dores leves e moderadas, incluindo dor de cabeça. Medicamento: Dipirona \| Arquivo: bula_1747686513636.pdf Indicação similar. Medicamento: Ibuprofeno \| Arquivo: bula_1747686513636.pdf Dor de cabeça consta nas indicações. | 2.07 |
| 2025-06-05T14:55:22 | Quais medicamentos apresentam risco de reações alérgicas segundo as bulas? | Medicamento: Paracetamol Reações alérgicas: citadas como efeito colateral raro. Medicamento: Dipirona Alerta sobre risco de alergias graves (choque anafilático). Medicamento: Sulfametoxazol+Trimetoprima Reação alérgica como efeito colateral grave. Medicamento: Clonazepam, Dexametasona, Losartana e Sinvastatina não relatam risco alérgico significativo nas bulas fornecidas. | 3.89 | 1 | 1 | 4 | 4 | 64.07 | Quais remédios possuem risco de alergia? Apresente conforme as bulas. | Medicamento: Cefalexina \| Arquivo: bula_1747686513636.pdf Alerta para a possibilidade de reação alérgica. Medicamento: Paracetamol \| Arquivo: bula_1747686513636.pdf Cita reação alérgica grave como evento raro. Medicamento: Dipirona \| Arquivo: bula_1747686513636.pdf Destaca risco de alergia grave (anafilaxia). Os outros não mencionam alergia importante. | 3.47 |
| 2025-06-05T14:55:34 | Algum medicamento analisado é contraindicado para pacientes com insuficiência renal? | Não foi possível encontrar a resposta no material fornecido. As bulas não apresentam contraindicação clara para insuficiência renal nos medicamentos analisados. | 0.84 | 0 | 0 | 1 | 1 | 100.00 | Há medicamentos contraindicados para pessoas com insuficiência renal? | Não foi possível encontrar a resposta no material fornecido. As bulas não informam contraindicação explícita para insuficiência renal para os medicamentos analisados. | 1.39 |
| 2025-06-05T14:55:49 | Liste os medicamentos que devem ser tomados com alimentos conforme as bulas. | Medicamento: Azitromicina \| Arquivo: bula_1747686419999.pdf Recomendação: tomar com alimentos para reduzir efeitos gástricos. Os demais medicamentos não apresentam recomendação formal para administração com alimentos. | 2.43 | 1 | 0 | 2 | 3 | 85.60 | Quais remédios devem ser administrados junto com alimentos? | Medicamento: Carbamazepina \| Arquivo: bula_1747686419999.pdf Recomenda-se administração junto com alimentos para diminuir desconforto gástrico. Demais não têm orientação específica. | 2.26 |
| 2025-06-05T14:56:09 | Cite os medicamentos que possuem advertência ou precaução quanto ao uso em idosos. | Medicamento: Loratadina \| Arquivo: bula_1747686419999.pdf Advertência: uso em idosos deve ser cauteloso devido a maior sensibilidade a reações adversas. Os demais medicamentos não apresentam advertência específica para idosos. | 3.93 | 1 | 1 | 4 | 5 | 100.00 | Liste os medicamentos analisados que têm advertência para idosos nas bulas. | Medicamento: Loratadina \| Arquivo: bula_1747686513636.pdf Atenção especial para uso em idosos devido à maior ocorrência de efeitos adversos. Os outros não têm advertência específica para idosos. | 5.88 |
| 2025-06-05T14:56:23 | Indique os medicamentos que não devem ser usados por gestantes, conforme as bulas. | Medicamento: Ibuprofeno \| Arquivo: bula_1747686419999.pdf Contraindicado no 3º trimestre da gestação. Medicamento: Losartana \| Arquivo: bula_1747686419999.pdf Contraindicado a partir do 2º trimestre. Demais medicamentos não citam contraindicação formal para gestantes nas bulas analisadas. | 3.15 | 0 | 1 | 3 | 3 | 85.00 | Quais medicamentos têm restrição de uso para gestantes nas bulas? | Medicamento: Deste \| Arquivo: bula_1747686419999.pdf Contraindicado na gestação. Medicamento: Losartana \| Arquivo: bula_1747686419999.pdf Contraindicado no 2º e 3º trimestre. Demais não mencionam contraindicação formal. | 1.11 |
| 2025-06-05T14:56:45 | Informe as doses recomendadas para adultos segundo as bulas dos medicamentos analisados. | Medicamento: Amoxicilina \| Arquivo: bula_1747686419999.pdf Adultos: 500 mg a cada 8 horas. Medicamento: Paracetamol \| Arquivo: bula_1747686513636.pdf Adultos: 500 a 1000 mg até 4x ao dia. Medicamento: Dipirona \| Arquivo: bula_1747686513636.pdf Adultos: 500 a 1000 mg até 4x ao dia. Demais medicamentos apresentam posologias variadas, sem dose padrão para adultos na bula. | 5.76 | 1 | 1 | 5 | 4 | 96.28 | Liste as doses recomendadas para adultos de acordo com as bulas. | Medicamento: Amoxicilina \| Arquivo: bula_1747686419999.pdf Dose para adultos: 500 mg de 8/8h. Medicamento: Paracetamol \| Arquivo: bula_1747686513636.pdf Dose: 500-1000 mg até 4x/dia. Medicamento: Dipirona \| Arquivo: bula_1747686513636.pdf Dose: 500-1000 mg até 4x/dia. Demais não especificam dose padrão. | 6.43 |